# Flexoelectricity-stabilized ferroelectric phase with enhanced reliability in ultrathin La:HfO$_2$ films

**Authors:** Peijie Jiao[1], Hao Cheng[1], Jiayi Li[1], Hongying Chen[1], Zhiyu Liu[1], Zhongnan Xi[1], Wenjuan Ding[1], Xingyue Ma[1], Jian Wang[1], Ningchong Zheng[1], Yuefeng Nie[1], Yu Deng[1], Laurent Bellaiche[3], Yurong Yang[1]\*, Di Wu[1,2]\*

**Affiliations:**

[1]National Laboratory of Solid State Microstructures, Department of Materials Science and Engineering, Jiangsu Key Laboratory of Artificial Functional Materials, and Collaborative Innovation Center of Advanced Microstructures, Nanjing University, Nanjing 210093, China.

[2]School of Materials Science and Intelligent Engineering, Nanjing University, Suzhou 215163, China.

[3]Physics Department, Institute for Nanoscience and Engineering, University of Arkansas, Fayetteville, Arkansas 72701, USA

\*Corresponding author. Email: yangyr@nju.edu.cn, diwu@nju.edu.cn

**Abstract:** Doped HfO$_2$ thin films exhibit robust ferroelectric properties even for nanometric thicknesses, are compatible with current Si technology and thus have great potential for the revival of integrated ferroelectrics. Phase control and reliability are core issues for their applications. Here we show that, in (111)-oriented 5%La:HfO$_2$ (HLO) epitaxial thin films deposited on (La$_{0.3}$Sr$_{0.7}$)(Al$_{0.65}$Ta$_{0.35}$)O$_3$ substrates, the flexoelectric effect, arising from the strain gradient along the film's normal, induces a rhombohedral distortion in the otherwise $Pca2_1$ orthorhombic structure. Density functional calculations reveal that the distorted structure is indeed more stable than the pure $Pca2_1$ structure, when applying an electric field mimicking the flexoelectric field. This rhombohedral distortion greatly improves the fatigue endurance of HLO thin films by further stabilizing the metastable ferroelectric phase against the transition to the thermodynamically stable non-polar monoclinic phase during repetitive cycling. Our results demonstrate that the flexoelectric effect, though negligibly weak in bulk, is crucial to optimize the structure and properties of doped HfO$_2$ thin films with nanometric thicknesses for integrated ferroelectric applications.

**Keywords**: HfO$_2$-based film, flexoelectric effect, ferroelectric properties, pulsed laser deposition, phase transition

**Main Text:**

Ferroelectricity of doped $HfO_2$ thin films in simple fluorite structure have attracted much attention since its first discovery in 2011 [1]. Robust ferroelectric properties observed in nanometer-thick thin films and the compatibility of $HfO_2$ with the complementary metal-oxide-semiconductor processes have tremendous potential to circumvent the challenges of integrated ferroelectrics that have difficulty combining perovskite ferroelectric thin films with state-of-the-art high-density semiconductor devices [2–5]. Various device concepts have been proposed to utilize ferroelectric properties in doped $HfO_2$ thin films, including ferroelectric random access memories [6], ferroelectric field effect transistors [5,7], negative capacitance transistors [8], ferroelectric tunnel junctions [9,10] and high-electron-mobility transistors [11].

Since the seminal work of Si-doped $HfO_2$ [1], ferroelectricity has been achieved in $HfO_2$ thin films, deposited using various techniques [12–14], by doping various elements [15–18]. Polycrystalline ferroelectric doped $HfO_2$ films, composed of centrosymmetric tetragonal (*t*-) and monoclinic (*m*-) phases companied with polar orthorhombic (*o*-) phase crystallites [19,20], have been obtained by annealing atomic-layer-deposited films sandwiched between top and bottom electrodes. On the other hand, epitaxial doped $HfO_2$ ferroelectric films have been achieved on yttria-stabilized zirconia [21], GaN [22], Si [23] and various perovskite oxide substrates [12,15,24]. Although epitaxial films have fewer defects and more controllable microstructures, structural control on these nanometer-thick thin films is still tricky due to the profound competition and subtle balance among t-, m- and o-phase polymorphs as functions of temperature [12], strain [24] and interfacial microstructures [25,26]. Although doped $HfO_2$ ferroelectric thin films are often assigned to the $Pca2_1$ o-phase, $Pnm2_1$ and $Pbca$ o-phases have also been reported [27,28]. Furthermore, a high energy rhombohedral (*r*-) phase with large polarization was reported for epitaxial $Hf_{0.5}Zr_{0.5}O_2$ (HZO) thin films [29]. How this *r*-phase is stabilized is still an enigma. It is essential to understand the microstructure of doped $HfO_2$ thin films under a variety of chemical, mechanical and electrostatic boundary conditions to optimize ferroelectric properties.

From the application point of view, fatigue is the most significant reliability issue for ferroelectric devices, especially for memories [15,28,30–32]. Unfortunately, doped $HfO_2$ thin films, as represented by intensively studied HZO films, fatigue upon repetitive bipolar switching [23,33]. For example, Lyu et. al. reported that the polarization of HZO thin films decreases from 16.0 to about 8.5 $\mu C/cm^2$ by only $10^6$ switches, and further to 2.3 $\mu C/cm^2$ by $10^9$ switches [23]. Following fatigue models proposed for conventional perovskite ferroelectrics, the fatigue in doped $HfO_2$ ferroelectric films has been ascribed to defect generation and/or charge injection that pins domain wall moving [28,30–32]. However, the metastable nature of the ferroelectric phase in doped $HfO_2$ thin films has been mostly overlooked so far. A detailed study on the fatigue mechanism and a proposal to enhance fatigue performance are strongly demanded for ferroelectric device applications of doped $HfO_2$ films.

Here we deposit 5%La:$HfO_2$ (HLO) epitaxial thin films with various thickness on (110)-oriented $(La_{0.3}Sr_{0.7})(Al_{0.65}Ta_{0.35})O_3$ (LSAT) substrates buffered with 20.0 nm thick $La_{0.67}Sr_{0.33}MnO_3$ (LSMO). La-doping into $HfO_2$ has been reported to greatly suppress the defect density, leading to a reduced leakage current [34,35] and an improved fatigue endurance [36]. With the help of density functional calculations, structure characterizations on the deposited and exfoliated HLO films, in comparison with HZO films, reveal that a *r*-distortion appears in HLO films thicker than 8.0 nm due to a strain-gradient-induced flexoelectric field. In contrast to conventional fatigue mechanisms reported in the literature [28,30–32], fatigue in doped $HfO_2$ thin films is found to occur as a transition from the metastable *o*-phase to the thermodynamically stable non-polar *m*-phase upon repetitive switching. The *r*-distortion observed in HLO films further stabilizes the metastable ferroelectric structure and dramatically suppresses fatigue.

**Structure evolution with increasing thickness**

It has been well established that strain, either from thermal coefficient mismatch or lattice mismatch [24], is essential to stabilize the metastable polar phases in various doped ferroelectric $HfO_2$ thin films. We first study the structure evolution with increasing HLO film thickness, as the strain imposed by the substrate relaxes with increasing thickness. Figure 1(a) shows X-ray diffraction (XRD) θ-2θ patterns of HLO/LSMO heterostructures deposited on (110) LAST substrates with increasing HLO thickness. Main feature in these XRD patterns is a peak at around 2θ=30°, usually indexed as the (111) plane of *o*-phase $HfO_2$, which is the metastable ferroelectric phase formed during cooling from the thermodynamically stable high temperature *t*-phase. Hereafter, *o*-phase and *m*-phase HLO is abbreviated as *o*-HLO and *m*-HLO, respectively. The *o*-HLO films deposited on (110) LSAT substrates is obviously (111)-oriented. (111)-oriented Zr- and Y-doped $HfO_2$ films have been previously reported on both (001) and (110) $SrTiO_3$ substrates [12,29,37], implying that it is not the lattice mismatch but probably the surface energy that dominates the oriented growth of doped $HfO_2$ thin films [38]. The clear Kiessig fringes observed in films thicker than 8.0 nm evidence the smooth surface and abrupt interface in these heterostructures. Another observation, with increasing HLO thickness, is that the (111) peak from *o*-HLO shifts obviously to lower 2θ angles, indicating the expansion of lattice spacing in the out-of-plane direction. In hetero-epitaxy, this is usually accompanied with the relaxation of in-plane tensile strain in the epitaxial layers [39], here, the *o*-HLO films. To obtain the average lattice spacing of the HLO film as a function of the distance from the HLO/LSMO interface, the top 6.0, 9.0 and 12.0 nm of the same 18.0 nm thick HLO sample was etched off and the crystal structure of the remaining films was checked by XRD. Figure S1(a) shows that the (111) peak of the *o*-HLO film shifts to lower 2θ angles with increasing remaining thickness, which indicates that the lattice spacing in the out-of-plane direction increases as the distance from the HLO/LSMO interface increases.

Phase pure *o*-HLO films can be obtained up to 15.0 nm. Above this thickness, an additional peak appears at around 28.3°, which can be assigned to the ($\bar{1}$11) diffraction of m-HLO. To determine where the m-HLO nucleates in thick films, the top 20.0 nm of a 30.0 nm thick HLO sample was etched off and the remaining 10.0 nm thick HLO layer was then checked by XRD. The ($\bar{1}$11) diffraction of *m*-HLO disappeared completely in the etched sample, as shown in Fig. S1(b), indicating that the thermodynamically stable *m*-HLO appears near the surface of thick HLO films, where, away from the interface, the HLO lattice is less strained. There is also observed a clear growth mode transition from a well-aligned 2-dimensional growth to a poorly-aligned 3-dimensional mosaic growth with increasing thickness (Fig. S2). The 3-dimensional growth usually appears to release the elastic energy due to the epitaxial strain [40,41]. Therefore, a significant strain gradient may occur in epitaxy HLO films as the strain from substrates relaxes within only a few tens of nanometers.

Figure 1(b) shows the XRD φ-scan of the ($\bar{1}$11) diffraction of the 20.0 nm thick *o*-HLO film at χ=71°. Like *o*-HZO films deposited on (110) $SrTiO_3$ substrate reported previously [37], six diffraction peaks appear as a result of two domains rotated from each other by 180° with respect to the out-of-plane [111] direction, with each domain contributing to three poles, namely ($\bar{1}$11), (1$\bar{1}$1) and (11$\bar{1}$), separated by φ=120°. Figures 1(c) and (d) show the XRD θ-2θ patterns of the six {111} diffractions at χ=71° with their reciprocal vectors having in-plane components compared with that of the (111) peak with its reciprocal vector being completely out-of-plane, for the 5.0 and 20.0 nm thick HLO films, respectively. The {111} peaks of the 5.0 nm thick HLO film all appear at exactly the same Bragg angle as the out-of-plane (111) diffraction, indicating $d_{111}=d_{\bar{1}11}=d_{1\bar{1}1}=d_{11\bar{1}}=2.946$ Å, corresponding to the $o$-phase with the eight <111> reciprocal vectors at the same length. However, in the 20.0 nm thick HLO film, the six peaks from the inclined {111} planes appear at a Bragg angle obviously higher than that of the (111) plane, giving rise to a lattice spacing $d_{\bar{1}11}=d_{1\bar{1}1}=d_{11\bar{1}}=2.942$ Å that is significantly smaller than that of the out-of-plane $d_{111}=2.985$ Å, indicating a clear $r$-distortion [12]. Wei $et.$ $al.$ has reported an epitaxial $r$-HZO ferroelectric films deposited on (001) LSMO/STO substrates [29]. Most recently, Yun $et.$ $al.$ revealed that $r$-phase structure is energetically unfavorable but 5%Y:HfO$_2$ epitaxial thin films could be in the $Pca2_1$ $o$-phase with a small $r$-distortion [12]. To clarify the effect of doping elements on the structure, we compare the structure of HLO and HZO thin films, both 10.0 nm in thickness, deposited on (110) LSMO/LSAT substrates. As shown in Fig. S3, although the HLO film shows a clear $r$-distortion with elongated $d_{111}$, the HZO film is in the pure $o$-phase in absence of any distortion with the eight {111} lattice spacings at the same value. Aliovalent (Y or La) substitution into the HfO$_2$ lattice is indeed more favorable to induce the $r$-distortion than the Zr substitution.

In order to quantify the structure evolution of HLO with increasing thickness, we measured the XRD θ-2θ diffractions of the {002} planes at χ=55° as a function of HLO thickness. As shown in Fig. 1(e), there is a broad diffraction peak around 2θ=35° for each HLO film, which can be deconvoluted into two components. Overall, the diffraction peak at lower 2θ values has about half the integrated intensity of the higher angle peak, indicating that one lattice constant (**a**) is longer than the other two (**b** and **c**) having similar values because the structural factors of the three {002} planes are close due to the nearly cubic structure. The structure with longer **a** and similar **b** and **c** constants is consistent with the $Pca2_1$ $o$-phase of HfO$_2$ [42–44]. It is observed that both peaks shift to higher angles with increasing HLO thickness. The corresponding lattice constants, **a** and **b**≈**c**, decrease with increasing HLO thickness, as plotted in Fig. 1f. This indicates again that the HLO films are subject to an in-plane tensile strain, which relaxes as the film thickness increases. As discussed above, the out-of-plane (111) lattice spacing increases with increasing HLO thickness, as also plotted in Fig. 1(f). This cannot be accommodated in the framework of an orthorhombic structure with α=β=γ=90°. However, the increased $d_{111}$ in thicker HLO films can be understood by introducing a small $r$-distortion [12], $i.e.$, keeping the interaxial angles α=β=γ=90°-δ with δ increasing with the increase of HLO thickness. This can be illustrated as stretching the orthorhombic unit cell along the [111] direction, as schematically shown in Fig. 1(g). Moreover, Figure 1(h) shows that the distortion angle δ, derived from the data in Fig. 1(f), indeed increases continuously from 0.38° for the 8.0 nm to 0.75° for the 20.0 nm thick HLO films.

To further investigate the crystal structure of these epitaxial HLO thin films, high-angle angular dark field scanning transmission electron microscopy (HAADF-STEM) measurements were carried out. Figure 2(a) shows a cross-sectional HAADF-STEM image of the 8.0 nm thick HLO film deposited on LSMO electrodes observed along the [1$\bar{1}$0] zone axis of LSAT. The clear Z-contrast indicates a sharp HLO/LSMO interface. There are two sets of atomic planes, one parallel to the film surface with a lattice spacing about 2.975 Å, and the other 71° away with a lattice spacing about 2.954 Å, in agreement with the (111) and ($\bar{1}$11) planes, respectively, in $o$-HLO with $r$-distortion. The inset shows the fast Fourier transform (FFT) within the red square, indicating the out-of-plane [111] direction, the [$\bar{1}$11] direction 71° away and the in-plane [$\bar{2}$11] direction parallel to the [1$\bar{1}$0] direction of LSMO. Estandía $et.$ $al.$ has proposed a domain matching epitaxy mechanism for HZO thin films deposited on (001) LSMO/DyScO$_3$ substrates [45], where the large lattice mismatch (which is impossible to maintain for conventional epitaxy) can be effectively reduced by forming blocks that matches better. Recently, this unusual epitaxy was also reported in HZO films deposited on (110) LSMO/SrTiO$_3$ substrates [37]. It is reasonable that this mechanism also works for epitaxial HLO films deposited on (110) LSMO/LSAT substrates. From the clear HLO/LSMO interface shown in Fig. 2(a), the domain epitaxy relation can indeed be identified, *i.e.* 6 *o*-HLO ($\bar{1}$11) planes matching 7 LSMO (1$\bar{1}$0) planes.

Figure 2(b) shows a cross-sectional HAADF-STEM image of the 30.0 nm thick HLO film deposited on LSMO electrodes observed along the [1$\bar{1}$0] zone axis of LSAT. Although the image shows apparently continuous lattice planes, close inspection reveals different lattice structures along the film normal. Close to the HLO/LSMO interface, the lattice is similar to that shown in Fig. 2a, exhibiting two atomic planes with lattice spacing about 2.983 and 2.946 Å, respectively. The FFT of the red square shown in the lower inset in Fig. 2(b) indicates that these two atomic planes are 71° away from each other, corresponding to the (111) and ($\bar{1}$11) planes of the *o*-HLO with *r*-distortion. Close to the film surface, the lattice is also composed of two sets of atomic planes. From the FFT of the yellow square shown in the upper inset in Fig. 2(b), one can identify that one of these two atomic planes is parallel to the film surface, but the other is 74° away, much larger than the angle in the *o*-HLO lattice close to the HLO/LSMO interface. Two atomic planes, 74° away from each other with the lattice spacing about 3.153 and 2.852 Å, are in good agreement with (111) and ($\bar{1}$11) planes of non-polar *m*-phase of HfO$_2$ [46]. The STEM images are consistent with the XRD results, clearly showing that the 8.0 nm thick HLO film is in a ferroelectric *o*-phase with *r*-distortion, and that the thermodynamically stable non-polar *m*-phase appears in thicker HLO films close to the film surface due to strain relaxation.

**Flexoelectricity stabilized rhombohedral distortion**

To elucidate the effect of the tensile epitaxial strain on the structure, the LSMO electrode was selectively etched off and the 15.0 nm thick freestanding HLO film was transferred to a platinized Si substrate, as shown schematically in Fig. 3(a). The transferred HLO film is expected to be free from any strain or strain gradient. Figure 3(b) show XRD θ-2θ patterns of the HLO/LSMO/LSAT heterostructure and the transferred HLO film on Pt/Si. The (111) diffraction of the transferred HLO film shifts to a slightly higher Bragg angle, compared to that of the epitaxial heterostructure, indicating that the out-of-plane (111) lattice spacing is reduced from 2.980 Å before to 2.962 Å after the exfoliation. It has been revealed, when discussing the structure evolution with HLO thickness, that the relaxation of in-plane tensile epitaxial strain results in the increase of out-of-plane lattice spacing accompanied by stronger *r*-distortion. The reduced out-of-plane lattice spacing observed in the transferred HLO film, in which the in-plane tensile epitaxial strain is completely released, cannot be explained simply on strain relaxation. Figures 3(c) and (d) show XRD θ-2θ patterns of the {111} diffractions before and after the exfoliation, respectively. This 15.0 nm thick HLO film on the LSMO/LSAT substrate shows a clear *r*-distortion with the (111) lattice spacing (about 2.980 Å) longer than the others (about 2.946 Å), while the same HLO film is in the pure *o*-phase in absence of any *r*-distortion after exfoliation, as evidenced by the same Bragg angle of all the {111} diffractions. It is the disappearance of the *r*-distortion that reduces the out-of-plane (111) lattice spacing in the exfoliated HLO film. This *r*-distortion observed in epitaxial HLO films requires both strain and strain relaxation, and it thus must be associated with the strain gradient and the corresponding flexoelectric effect [47,48]. Ge *et. al.* recently also reported that the (111)-oriented HZO thin films in the apparent *r*-phase transits into the *o*-phase after exfoliated from the (001) SrTiO$_3$ substrate [49].

A direct consequence of the flexoelectric field in ferroelectric thin films is the appearance of imprint, as a horizontal asymmetry in the *P-E* loops. Our epitaxial HLO thin films are ferroelectric as-deposited. As shown in Fig. 4(a) and (b), the *P-E* loop of the 8.0 nm thick HLO is almost symmetric, while that of the 15.0 nm thick HLO shifts clearly to the right, indicating the existence of an internal electric field pointing upward. This also suggests that the observed internal field is associated with strain relaxation [50]. Figure 4(c) shows the amplitude of the internal field, represented by the algebraic sum of the positive and negative coercive fields, as a function of film thickness. The 10.0 and 15.0 nm thick HLO films, both with considerable *r*-distortion, show significant positive imprints, about 0.41 and 0.54 MV/cm, respectively, while the 8.0 nm thick HLO, 10.0 nm thick HZO and exfoliated 15.0 nm thick HLO films all show small negative imprints. Corresponding *P-E* loops are shown in Fig. S4. The much weaker internal electric field in the HZO film is consistent with the structural analysis that HZO is more resistant to the *r*-distortion. Compared with the HZO film, the HLO film in the same thickness shows a stronger internal field that induces the large observed *r*-distortion. The almost symmetric *P-E* loops of the transferred 15.0 nm thick HLO film, in pure *o*-phase, clearly correlate the internal field with strain and strain relaxation. This internal field, leading to the observed imprint, is the flexoelectric field due to the strain gradient.

To further investigate the stability and effects of the *r*-distortion, density functional theory (DFT) calculations were performed on both 6.25%La-doped and 50%Zr-doped $HfO_2$. The compositions were chosen to best mimic the experimental ones within our calculations' capability. We start from the *Pca*$2_1$ *o*-phase where the interaxial angles are all 90°. The *r*-distortion is introduced by reducing the interaxial angles by δ and keeping α=β=γ=90°-δ. To simulate the effect of flexoelectricity, an electric field of 0.05 or 0.08 V/Å is applied along the [111] direction in our calculations [51,52]. The electric enthalpy *F* is calculated as *F=U-**P·E*** in the unit volume, where *U* is the Kohn-Sham energy, ***P*** the polarization, and ***E*** the applied electric field.

Figure 4(d) displays the electric enthalpy as a function of the *r*-distortion angle δ under the effective flexoelectric field. With the 0.05 V/Å electric field, although the minimum electric enthalpy of HZO is still at about δ=0° in the pure *o*-phase, that of HLO appears already at about δ=0.2°, *i.e.*, with a *r*-distortion. As the electric field increased to 0.08 V/Å, both HZO and HLO exhibit a minimum electric enthalpy at a finite *r*-distortion angle δ, that is about 0.1° for HZO and about 0.3° for HLO. In other words, the *o*-phase ferroelectric doped $HfO_2$ tends to exhibit a *r*-distortion under a stronger enough flexoelectric field, rather than keeping the perfect *o*-phase. Compared with HZO, HLO gains more energy and is it thus easier for it to exhibit the *r*-distortion. This is consistent with the experimental results that the 10.0 nm thick HLO film shows the *r*-distortion, while the 10.0 nm thick HZO film does not.

**Fatigue characterizations of HLO films**

Figure 5(a) shows bipolar fatigue characteristics of HLO films, 8.0 and 10.0 nm in thickness, at room temperature. The fatigue characteristics of the 10.0 nm thick HZO film is also shown for comparison. The *P-E* loops of these films as functions of switching cycles are shown in Fig. 5. The amplitudes of the triangular cycling pulses are 5.5 V for the 8.0 nm thick HLO, 7.0 V for the 10.0 nm thick HLO and 5.0 V for the 10.0 nm thick HZO films, respectively, more than 2 times of their respective coercive fields and sufficient to saturate the polarization. The 8.0 nm thick HLO film fatigues severely with its $P_r$ dropping abruptly by 36.0% at about only $10^4$ switching cycles. Pešić *et. al*. reported that the main mechanism responsible for the degradation of the ferroelectric behavior is domain pinning and trap generation [30]. Cheng *et. al*. reported a fatigue/rejuvenate process associated with a reversible polar $Pbc2_1$ and antipolar $Pbca$ phase transition in polycrystalline HZO thin films [28]. However, the 10.0 nm thick HLO film shows much better fatigue endurance up to $10^9$ cycles with the reduction of $P_r$ less than 11%. In comparison, $P_r$ of the 10.0 nm thick HZO film decreases continuously with repetitive switching and only keeps 62% of its initial value after $10^9$ cycles.

To elucidate the fatigue mechanism, HADDF-STEM analysis was performed on the cross-section of the 8.0 nm thick HLO capacitor fatigued after $10^4$ switching cycles. As shown in Fig. 5(b), $m$-HLO crystallites, as indicated by the 3.153 Å lattice spacing for its ($\bar{1}11$) plane parallel to the film surface, appear at the surface of this fatigued 8.0 nm thick HLO. The inset is a zoom-in of the red square on the image, showing two atomic planes, 74° away from each other, consistent with $m$-HLO [53]. The appearance of $m$-phase in the fatigued HLO film, which is in absence of this non-polar phase before cycling, strongly suggests that fatigue in doped $HfO_2$ occurs with the transition from metastable ferroelectric $o$-phase to thermodynamically stable but non-polar $m$-phase under repetitive switching. The polarization decreases with the reduction of volume that can be switched. In this way, the much better fatigue performance observed for the 10.0 nm thick HLO film can be understood because the flexoelectric field induced $r$-distortion further stabilizes the ferroelectric $o$-phase, making it more robust against the transition to the non-polar $m$-phase. On the contrary, the 8.0 nm thick HLO film is close to the $o$-phase with a very small $r$-distortion and the 10.0 nm thick HZO film is in the pure $o$-phase without the $r$-distortion. They are more vulnerable to the transition from $o$- to $m$-phase, which leads to severe fatigue.

Figure 5(c) shows the fatigue characteristics of the 15.0 nm thick HLO films measured with 7.5, 8.5 and 9.5 V bipolar pulses. The 15.0 nm thick HLO films shows an even better fatigue performance since $P_r$ decreases by only 7% after $10^9$ switches under 7.5 V cycling, as also represented by the identical $P$-$E$ loops measured during cycling shown in Fig. 5(d). However, it is obvious that stronger pulses induce larger polarization loss, about 20% and 40% reduction, respectively, under 8.5 and 9.5 V cycling up to $10^9$ switches. Corresponding hysteresis loops as functions of switches are shown in fig. 6. This is consistent with the above proposed fatigue mechanism. The simplest scenario could be a temperature-driven transition [54,55]. Stronger pulses, which inject more energy into the capacitor, may increase the local temperature more quickly and hence drive more transition from $o$- to $m$-phase.

**Summary**

The structure and ferroelectric properties of (111)-oriented HLO films, epitaxially deposited on (110) LSMO/LSAT substrates, have been studied. Strain relaxation was observed with the increase of film thickness in HLO thin film thicker than 8.0 nm. The strain gradient along the film normal results in a flexoelectric field, which in turn induces a $r$-distortion in the otherwise $Pca2_1$ $o$-phase. Compared with doped $HfO_2$ thin films in pure $o$-phase, this $r$-distortion further stabilizes the metastable ferroelectric $o$-phase by reducing its enthalpy with the existence of the flexoelectric field. Therefore, the HLO films with this $r$-distortion show much better fatigue resistance, by keeping more than 92% of its initial $P_r$ after $10^9$ bipolar switching cycles, than those in pure $o$-phase, which fatigue severely by transition of the metastable ferroelectric $o$-phase into non-polar $m$-phase under repetitive switching. Our work demonstrates that the flexoelectric effect, negligibly weak in bulk, can be an important factor in optimizing the structure and properties of nanoscale ferroelectrics and the fatigue performance of doped $HfO_2$ can be enhanced by further stabilizing the metastable ferroelectric phase.

**Methods section**

*Sample deposition*: The LSMO and doped HfO$_2$ (HLO and HZO) layers were deposited sequentially on (110)-oriented LSAT single crystalline substrates by pulsed laser deposition (AdNano Corp.) both at 750 under 100 mTorr oxygen pressure. 50%Zr:HfO$_2$, 5%La:HfO$_2$ and LSMO polycrystalline targets were all synthesized by conventional solid state reaction. The ablation was performed with a pulsed 248 nm output from a PLD20 KrF excimer laser (Excimer, China) with an energy density of 1.5 J/cm$^2$ on the targets and a repetition rate of 2 Hz. Deposition rate of the HZO, HLO and LSMO films was calibrated as 0.006, 0.005 and 0.007 nm/pulse, respectively, by fitting the X-ray reflection data collected using a Brucker D8 Discover diffractometer. In this work, the LSMO electrode is kept at 20.0 nm in thickness, while the HLO film varies from 2.5 to 37.5 nm in thickness. The control HZO thin film is 10.0 nm in thickness.

*Structure characterizations*: The crystal structure of the HZO, HLO and LSMO layers were examined by XRD using the same Brucker D8 Discover diffractometer. Aberration-corrected HAADF-STEM imaging of the heterostructure cross-section was acquired by an FEI Titan3 G2 60-300 microscope at 300 kV. The STEM samples were prepared using a Helios G4 UX (Thermo Fisher) focused ion beam setup. Surface morphology of the HLO films was characterized using an Asylum Research Cypher ES atomic force microscope (AFM). Root-mean-square roughness was acquired over a 3×3 μm$^2$ area. Dry etching of the HLO film was achieved using a GSE 200 inductively coupled plasma etching system (NAURA Technology Group Co. Ltd.). Selective wet etching of the LSMO electrode to obtain freestanding HLO films was performed by slowly immersing the films with substrate in a beaker of aqueous solution (200 mL) of KI (4 mg) and HCl (5 mL), followed by keeping the beaker in a 40 °C water bath for several hours. The released HLO films floating on the water surface was then collected by a platinized Si substrate and placed on a 75 °C hot plate to remove the deionized water.

*r-distortion angle determination*: The *r*-distortion was introduced by reducing the interaxial angles while keeping them equal, *i.e.*, α=β=γ=90°-δ. The d$_{111}$ lattice spacing can be expressed with lattice constants **a**, **b**, **c**, and the interaxial angle α as,[56]

$$\frac{1}{d_{111}^2} = \frac{1}{V^2}(S_{11} + S_{22} + S_{33} + 2S_{12} + 2S_{23} + 2S_{13}), \tag{1}$$

where

$$V = abc\sqrt{1 - 3cos^2\alpha + 2cos^3\alpha} \tag{2}$$

$$S_{11} = b^2c^2 sin^2\alpha \tag{3}$$

$$S_{22} = a^2c^2 sin^2\alpha \tag{4}$$

$$S_{33} = a^2b^2 sin^2\alpha \tag{5}$$

$$S_{12} = abc^2(cos^2\alpha - cos\alpha) \tag{6}$$

$$S_{23} = a^2bc(cos^2\alpha - cos\alpha) \tag{7}$$

$$S_{13} = ab^2c(cos^2\alpha - cos\alpha). \tag{8}$$

Substituting equations (2)-(8) into equation (1) and applying **b≈c**, one achieves,

$$\frac{1}{d_{111}^2} = \frac{[(b^2 + 2a^2)sin^2\alpha + 2(a^2 + 2ab)(cos^2\alpha - cos\alpha)]}{a^2b^2(1 - 3cos^2\alpha + 2cos^3\alpha)} \tag{9}$$

The distortion angle δ can be determined by solving equation (9) with the **a**, **b** and $d_{111}$ data presented in Figure 1f.

*Ferroelectric characterizations*: *P-E* hysteresis loops and fatigue endurance characteristics were studied using a Radiant Precision Multiferroic system in virtual ground mode. Pt top electrodes, 30 μm in diameter, were deposited through a shadow mask using an AJA Orion-8-UHV sputtering system at room temperature. Ohmic contact to the LSMO electrode was achieved by silver paste. The LSMO electrode is grounded in all the measurements. Fatigue characteristics were measured by cycling under 100 kHz bipolar pulses of various amplitudes.

*Density functional theory calculations*: DFT calculations were performed via the Vienna Ab initio Simulation Package (VASP) [57]. The projector augmented wave method[58] combined with the PBEsol functional[59] was used in the electronic structure calculation with the cutoff energy as 550 eV. Hf $5s^25p^66s^25d^2$, Zr $4s^24p^65s^24d^2$, La $5s^25p^66s^25d$ and O $2s^22p^4$ were considered as valence electrons. The structures were relaxed until the force at each atom was less than 0.001 eV Å$^{-1}$. The equilibrium structure under electric field is achieved by minimizing the electric enthalpy[60,61]. The polarization was calculated by producing atomic displacements with respect to the paraelectric cubic phase and the Born effective charges[62]. A unit cell with 2 Hf, 2 Zr and 8 O atoms was used to simulate the HZO lattice where 50% Hf atoms in $HfO_2$ was substituted by Zr. A supercell composed of 2×2×2 $HfO_2$ unit cells with 96 atoms was used to simulate HLO, in which 2 Hf atoms were replaced by La and one oxygen vacancy was introduced to keep the charge neutrality (Figure S7). The composition of the supercell corresponds to 6.25%La-doped $HfO_2$, which is very close to the experimental composition used.

**Acknowledgments:**

**Funding:** This work was supported by Natural Science Foundation of China (grant nos. 52232001, 51725203, 51721001, 52003117, and 11874207), and the National Key R&D Program of China (grant no. 2020YFA0711504). L.B. thanks the Office of Naval Research (Grant No. N00014-21-1-2086) and the Vannevar Bush Faculty Fellowship (VBFF) Grant No. N00014-20-1-2834 from the Department of Defense. Y.D. thanks to the Fundamental Research Funds for the Central Universities No.021314380213 and Natural Science Foundation of Jiangsu Province, China (grant no BK20201246). Y.R.Y. thanks Dr. Massimiliano Stengel for valuable discussions on the flexoelectric effects.

**Author contributions:** D.W. conceived and directed the project with Y.R.Y.. P.J.J. deposited the films and characterized the structure and electrical properties with the help of H.Y.C., W.J.D. and N.C.Z. under the supervision of D.W. and Y.F.N.. J.Y.L. performed STEM measurements with the help of Z.Y.L. under the supervision of Y.D.. H.C. performed the density functional theory calculations with the help of X.Y.M. and J.W. under the supervision of Y.R.Y., L.B. and D.W.. D.W. and P.J.J. wrote the manuscript with input from Y.R.Y, Y.D. and Y.F.N.. All authors discussed the data and contributed to the manuscript.

**Competing interests:** Authors declare that they have no competing interests.

**Data and materials availability:** The data that support the plots within this paper and other findings of this study are available from the corresponding authors upon reasonable request.

**Supplementary Materials**

Figures S1 to S7

Supplementary Reference

**Figures**

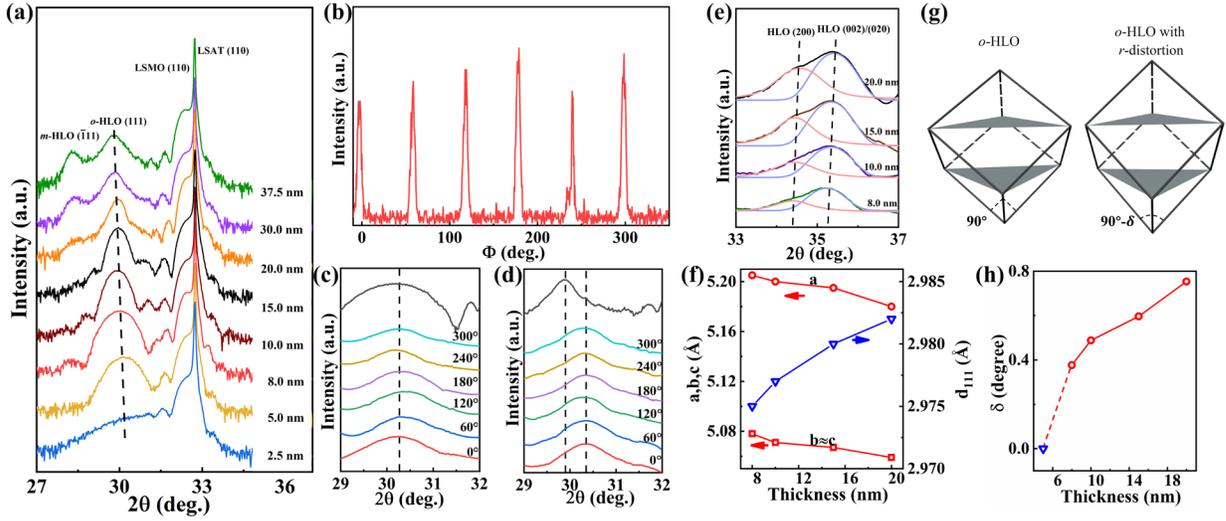

**Fig. 1. Structure evolution of HLO with film thickness.** (a) XRD θ-2θ patterns of HLO films in various thickness deposited on (110) LSMO/LSAT substrates. The dash line is a guide to the eyes. (b) XRD φ-scan pattern of *o*-HZO (111) reflections of the 20.0 nm thick HLO film shown in (a). XRD θ-2θ patterns of the {111} peaks of the (c) 5.0 and the (d) 20.0 nm thick HLO films, respectively, shown in (a). (e) XRD θ-2θ patterns of the {002} peaks of HLO films in various thickness. (f) The (100), (010), (001) and (111) lattice spacing as a function of HLO film thickness. (g) Sketches of a pure orthorhombic lattice and an orthorhombic lattice with a *r*-distortion. (h) The *r*-distortion, represented by the angle deviated from 90º, δ, as a function of HLO film thickness. δ=0 for the 5.0 nm thick HLO film in pure *o*-phase is also shown.

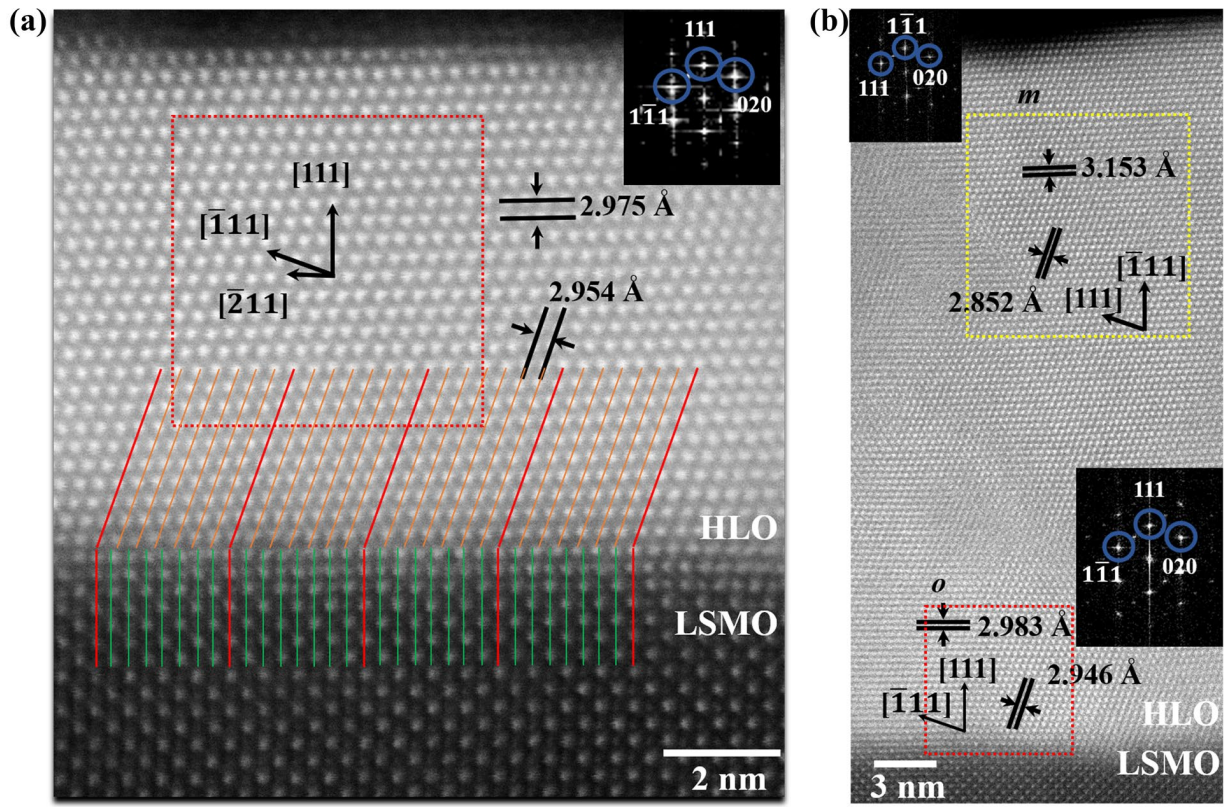

**Fig. 2. Cross-sectional TEM characterization of HLO films.** HAADF-STEM images of the (a) 8.0 and (b) 30.0 nm thick HLO films, respectively, observed along [001] zone axis of LSAT. The inset in (a) shows the FFT of the lattice in the red square. The upper (lower) inset in (b) shows the FFT of the lattice in the yellow (red) square.

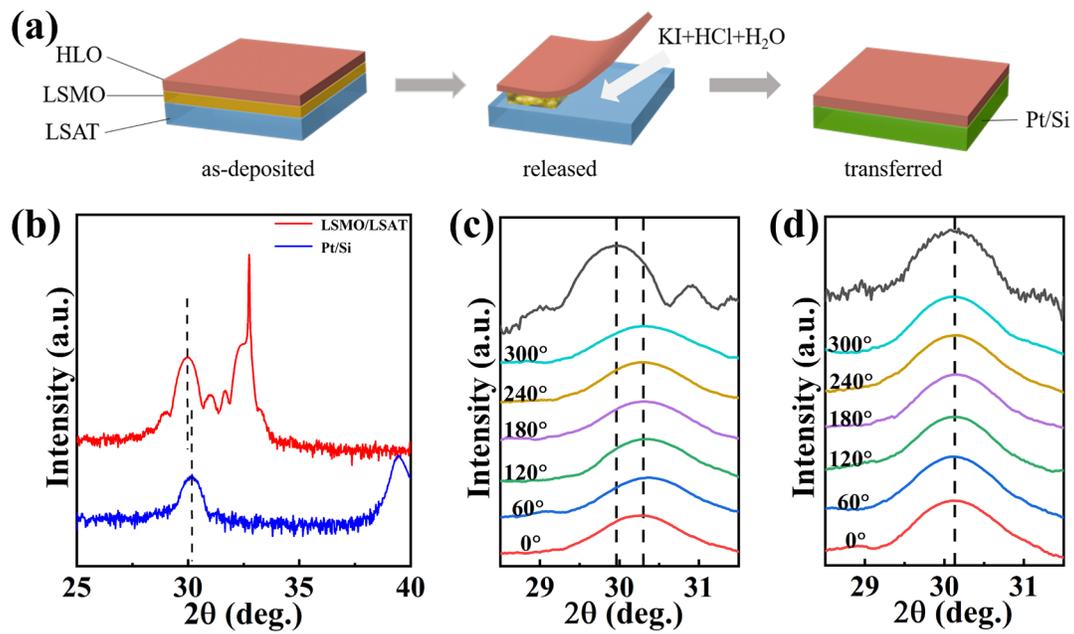

**Fig. 3. Structure evolution of HLO free from strain gradient.** (a) Schematic drawing of selective etching the sacrificial LSMO electrode layer in a diluted KI + HCl solution to achieve a freestanding HLO film. (b) θ-2θ patterns of the strained HLO film, 15.0 nm in thickness, on (110) LSMO/LSAT substrate and the same HLO film transferred on a Pt/Si substrate. XRD θ-2θ patterns of the {111} peaks of the (c) as-deposited and (d) transferred HLO films shown in (b).

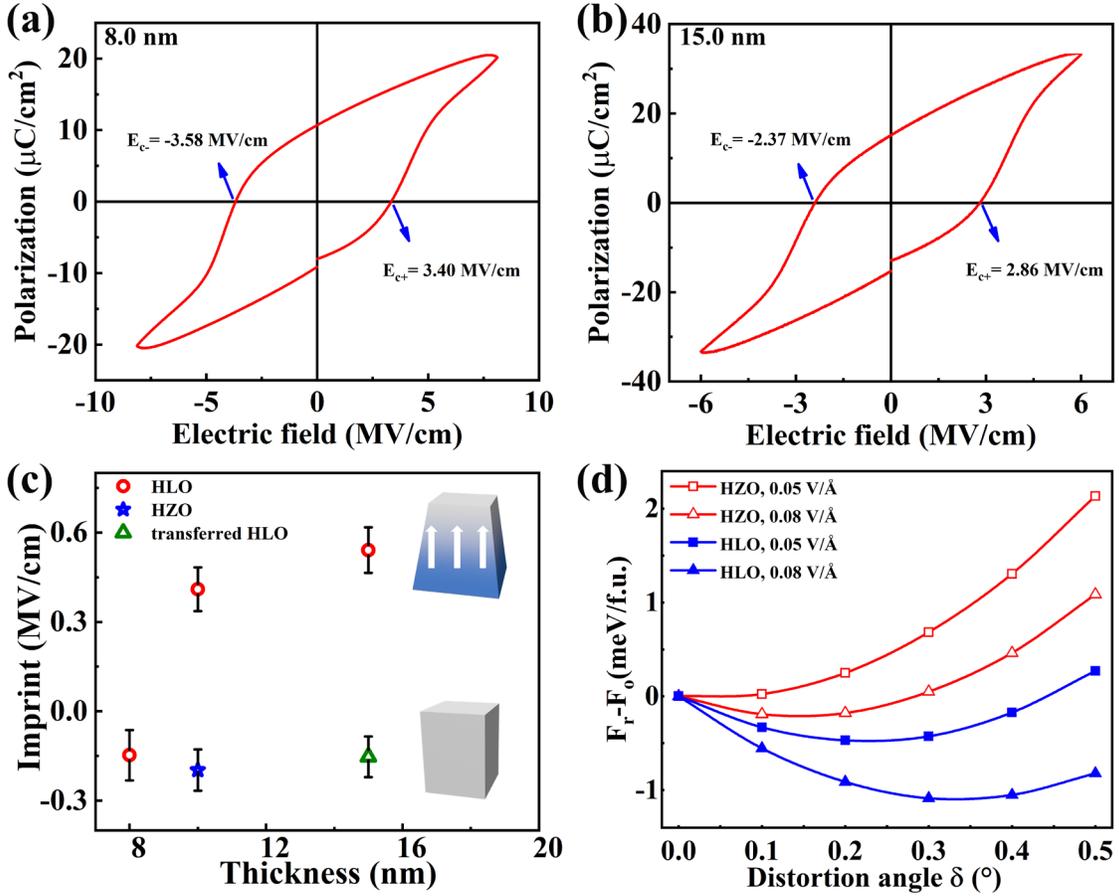

**Fig. 4. Flexoelectricity induced *r*-distortion.** Ferroelectric hysteresis loops of the (a) 8.0 and (b) 15.0 nm thick HLO films deposited on (110) LSMO/LSAT substrates, respectively. (c) Imprint of HLO films deposited on (110) LSMO/LSAT substrates as a function of film thickness. The imprints of a 10.0 nm thick HZO film deposited on (110) LSMO/LSAT substrate and a 15.0 nm thick HLO film transferred on Pt/Si substrate are also shown for comparison. The insets are sketches showing the imprint field due to a strain-gradient-induced flexoelectric effect. (d) Electric enthalpy of the rhombohedrally distorted *o*-phase as a function of *r*-distortion angle δ for HLO and HZO with different electric field applied along [111]. $F_o$ and $F_r$ are the electric enthalpy of perfect *o*-phase and distorted *o*-phase.

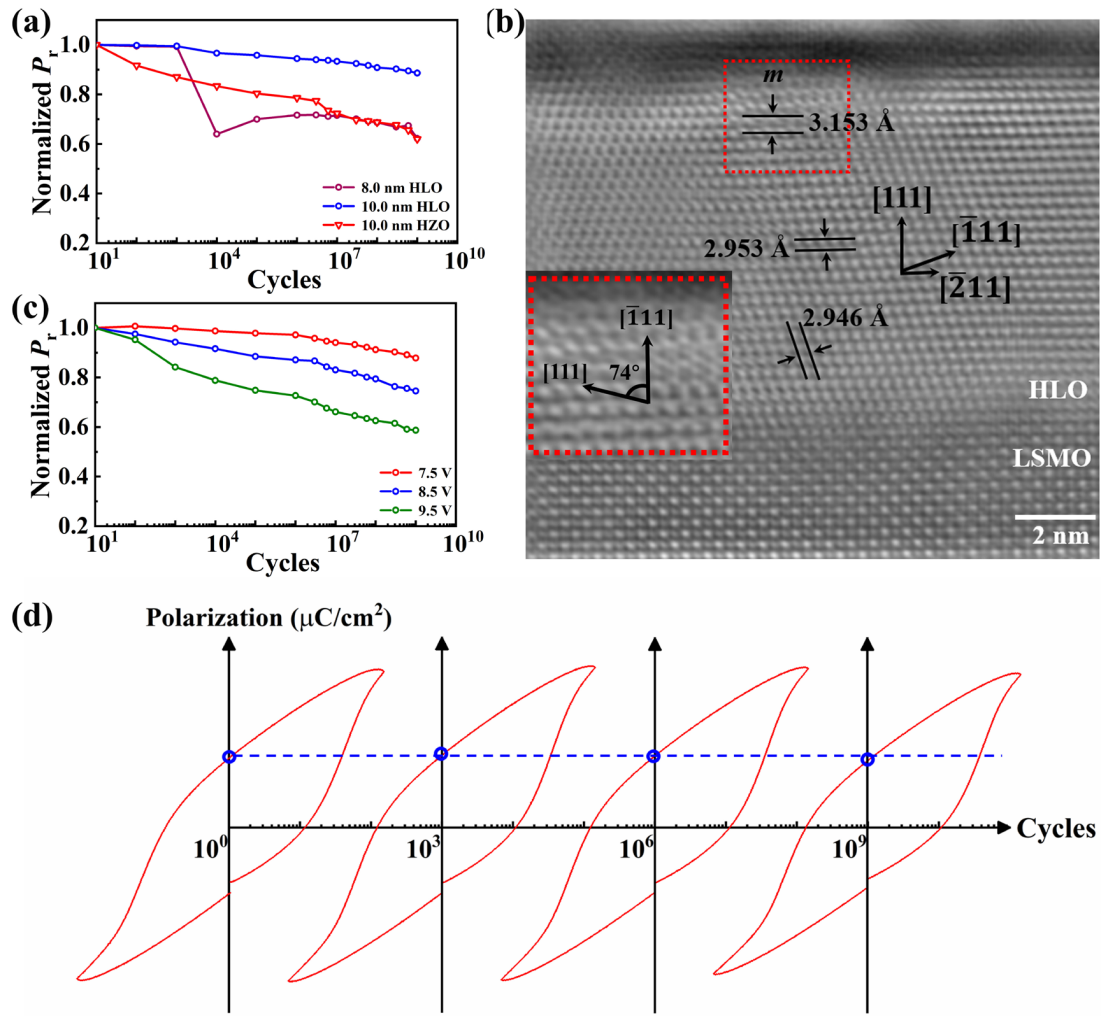

**Fig. 5. Fatigue characteristics of HLO films.** (a) Fatigue characteristics of HLO films in various thickness in comparison with a 10.0 nm thick HZO film; (b) Cross-sectional HAADF-STEM image of a fatigued 8.0 nm thick HLO film observed along the [001] zone axis of LSAT after $10^4$ switches. The inset is a zoom-in of the red square. (c) Fatigue characteristics of the 15.0 nm HLO film as functions of bipolar pulses in various amplitudes. (d) Ferroelectric hysteresis loops of the 15.0 nm thick HLO film as functions of 7.5 V switching cycles. The horizontal blue dash line indicates the original $P_r$.

24